## Query Details

1. Author names: Please confirm if the author names are presented accurately and in the correct sequence (given name, middle name/initial, family name). Author 1 Given name: [Pralay Kumar] Last name [Karmakar]. Also, kindly confirm the details in the metadata are correct.

2. Kindly check whether the inserted citations of Fig. 1 and Tables 2, 3 is correct and amend if necessary.

3. A data availability statement is mandatory for publication in this journal. Please provide the statement.

Review Paper

# Acoustic waves in the jovian dusty magnetosphere: a brief review and meta-analysis

Souvik Das    Affiliationids : Aff1

Ahmed Atteya    Affiliationids : Aff2

Pralay Kumar Karmakar ✉
Email : pkk@tezu.ernet.in
Affiliationids : Aff1, Correspondingaffiliationid : Aff1

Aff1 Department of Physics, Tezpur University, Napaam, Tezpur, Assam, 784028, India

Aff2 Department of Physics, Faculty of Science, Alexandria University, Alexandria, 21511, Egypt

## Abstract

The omnipresence of dust particulates in space and astrophysical plasmas has been attracting numerous researchers to study the collective excitation and AQ1 propagation dynamics of different eigen-mode structures in diversified astrocosmic circumstances for years. It includes planetary rings, interplanetary space, cometary tails, asteroid zones, planetary atmospheres, etc. The ubiquitous charged dust particulates possess collective degrees of freedom resulting in the excitation of relatively low-frequency modes, such as dust-ion-acoustic waves (DIAWs), dust-acoustic waves (DAWs), dust-Coulomb waves (DCWs), and so forth. An interesting prevalency of dusty plasma stability research lies in the Jupiter's magnetosphere (Jovian plasmas) embedded inside the supersonic solar wind. A brief review of the updated research works on dust-acoustic waves and related collective instability dynamics in the presence of trapped plasma particles is presented herein. The key aim of the proposed explorative meta-analysis is rooted in outlining concisely the main up-to-date investigations on such collective instability processes chronologically. An especial attention is given primarily to the thermostatistical description laws of the constitutive lighter electrons and ions against the heavier positively charged dust grains (microspheres). The trapping mechanism of both the lighter species (electrons + ions) is another additive feature revisited here properly. Finally, we clearly extrapolate a number of futuristic directions in light of sensible novelties with a wider scope.

## Keywords

Jupiter
Jovian magnetosphere
Jovian plasmas
Acoustic waves

# 1. Introduction

Since the space age has begun, human beings have been developing interest to study the plasma environments of the solar system and its constituent planets and moons. It is noteworthy that Jupiter, the fifth (in line from the Sun) and the largest planet in our solar system, has many mysterious astroplasmic facts of dynamical interest yet to be well investigated (Bolton et al. 2015; Gurnett et al. 2002; Saleem et al. 2012; Saur et al. 2018; Tsuchiya et al. 2018). Apart from it, there exist various interesting observational predictions of diversified structure formation in the Jovian dust ring systems, auroras, radiation belts, and so forth (Al-Yousef et al. 2021; Bolton et al. 2015; Connerney et al. 2017; Gurnett et al. 2002; Saleem et al. 2012; Saur et al. 2018; Smith et al. 1992; Tolba et al. 2021; Tsuchiya et al. 2018). It has made the Jovian magnetosphere one of the most captivating plasma environments yet to be well explored. It may be seen that, unlike the terrestrial planets, Jupiter does not have a solid rocky surface; instead, it is mainly composed of gaseous phases of Hydrogen (H) and Helium (He) (Bolton et al. 2015). This gas giant is known to have the largest planetary magnetosphere with the strongest planetary magnetic field ($\approx 8$ G $\sim 20$ times stronger than the Earth's magnetic field) in the solar system (Bolton et al. 2015; Connerney et al. 2017; Moslem et al. 2019; Tolba et al. 2021). It also possesses the strongest planetary radiation belt and the most powerful auroras (Bolton et al. 2015). The Jovian magnetosphere constituted of positively charged dust grains (Al-Yousef et al. 2021; El-Taibany et al. 2022; Moslem et al. 2019; Saleem et al. 2012; Tolba et al., 2017, 2021), is actually embedded within the supersonic solar wind plasma, interacting with the high energetic electrons and ions streaming from the Sun (Al-Yousef et al. 2021; Bolton et al. 2015; Connerney et al. 2017; Gurnett et al. 2002; Moslem et al. 2019; Saleem et al. 2012; Smith et al. 1992; Tsuchiya et al. 2018). As a result, the Jovian magnetosphere behaves as a unique complex plasma fluid medium exhibiting a plethora of collective excitation of waves, oscillations, and instabilities sourced in free microfluidic currents associated with its diversified constitutive species (Al-Yousef et al. 2021; Bolton et al. 2015; Connerney et al. 2017; El-Taibany et al. 2022; Gurnett et al. 2002; Saleem et al. 2012; Saur et al. 2018; Tolba et al. 2021). A spectrum of the diversified collective waves and instabilities excitable in the Jovian plasma environments alongside the corresponding sources is summarily highlighted in Table 1.

Table 1

A spectrum of Jovian plasma waves

| S No | Wave pattern | Physical mechanism | Evolution equation | Stability | Source |
|---|---|---|---|---|---|
| 1 | Solitons | Streaming instability saturation when nonlinear wave steepening (lowest-order) balances linear dispersive wave broadening (ignorable dissipation) | KdV & NLS equations | Stable (conservative) | Ahmad et al. (2013); Saleem et al. (2012); Tolba et al. (2017) |
| 2 | Rogue waves | Same as solitons, but becomes unstable by gaining energy from the random background smaller waves | Lowest order solution of NLS equation | Unstable (non-conservative) | Tolba et al. (2017) |
| 3 | Shocks | Streaming instability saturation when nonlinear wave steepening (lowest-order) balances linear wave dissipative decay (ignorable dispersion) | Burgers equation | Unstable (non-conservative) | Saleem et al. (2012) |
| 4 | Double layers | Higher-order nonlinear transformation and saturation of solitons | m-KdV equation | Unstable (non-conservative) | Crary and Bagenal (1997) |
| 5 | Vortices | Fast rotating planetary turbulence generates deep axially aligned cyclones and anticyclones swirling in strongly sheared deep zonal flow | QG equation | Unstable (rotatory) | Flierl et al. (2019); Yadav et al. (2020) |
| 6 | Jets | Deep zonal turbulent convection and stratospheric thermal perturbation | QG equation | Stable (conservative) | Dowling (2020); Flierl et al. (2019); Marcus (2004); Yadav et al. (2020) |
| 7 | Alfven waves | Strong electromagnetic coupling of Jovian strong magnetic field (magnetosphere) and the volcanically active Io's plasma torus (ionosphere) | MHD equations | Unstable (non-conservative) | Saur et al. (2018); Su et al. (2006) |
| 8 | Radio waves (Hectometric, Decametric) | Solar wind-created interplanetary shocks and Io-generated plasmas interact with Jovian magnetosphere leading to non-thermal electron–ion maser effects causing radio emissions | EM wave equation | Unstable (non-conservative) | Connerney et al. (2017); Gurnett et al. (2002); Su et al. (2006); Zarka (2004) |
| 9 | Whistler waves | Jovian lightning induces very low-frequency radio wave driven by electron beam with superthermal distribution | EM wave equation | Unstable (non-conservative) | Imai et al. (2018); Kolmašová et al. (2018); Leubner (1982) |
| 10 | Thermal waves | Due to magnetosphere-ionosphere coupling, Jovian tropospheric temperature shows global-scale longitudinal perturbations and variations which causes thermal waves | Heat flow equation | Stable (conservative) | Deming et al. (1997); Li et al. (2006) |
| 11 | Dus-thermal waves | Same as above but with fluctuations in dust thermodynamics | Heat flow equation | Stable (conservative) | Rao (2000) |
| 12 | Dust-ion-acoustic waves (DIAWs) | Longitudinal waves (low-frequency) excited by the rhythmic action of ion inertial force and plasma thermal pressure force with massive dust grains | Normal fluid equations (linear regime); KdVB, NLS, KP equations, etc. (nonlinear regime) | Unstable (non-conservative) | Ata-Ur-Rahman et al. (2019); Rao (1999); Shukla and Mamun (2001) |
| 13 | Dust-acoustic waves (DAWs) | Longitudinal waves (lower frequency) excited by the rhythmic action of dust inertial force and plasma thermal pressure force | Normal fluid equations (linear regime); KdVB, NLS, KP equations, etc. (nonlinear regime) | Unstable (non-conservative) | Al-Yousef et al. (2021); El-Taibany et al. (2022); Rao (1999); Shukla and Mamun (2001) |
| 14 | Dust-Coulomb waves (DCWs) | Grain charge fluctuations in dense dusty plasmas give rise to ultra-low-frequency electrostatic dust normal modes (inertia and thermal pressure by the dust grains) | Same as above but modified by dust fugacity effects | Unstable (non-conservative) | Rao (1999, 2000) |
| 15 | Dust lattice waves (DLWs) | In strongly coupled dusty plasma the electrostatic interaction energy of the shielded grains becomes much larger than the kinetic energy of the dust grains gives rises to dust lattice waves | KdV & NLS equation | Unstable (non-conservative) | Rao (1999, 2000) Shukla and Mamun (2001) |

| S No | Wave pattern | Physical mechanism | Evolution equation | Stability | Source |
|---|---|---|---|---|---|
| 16 | Cnoidal waves | The nonlinear interaction between the streaming electrons and ions of the solar wind with the Jupiter magnetosphere causes the dust-acoustic nonlinear periodic waves called the cnoidal wave | KdV equation | Unstable (non-conservative) | Tolba et al. (2021); Tolba et al. (2017) |
| 17 | Rossby waves | Naturally occurs in fast rotating Jovian plasma fluid due to conservation of potential vorticity (with a proper balance of inertia, buoyancy, pressure gradient, and Coriolis forces) | Rotational fluid equations | Unstable (non-conservative) | Allison (1990); Gu and Hsieh (2011); Theiss (2006); Warneford and Dellar (2017) |
| 18 | Rossby-gravity /Inertia-gravity /Yanai waves | Same as above but are more confined to the equatorial region | Rotational fluid equations | Unstable (non-conservative) | Allison (1990); Gu and Hsieh (2011); Orton et al. (2020); Warneford and Dellar (2017) |
| 19 | Tidal waves | Gravitational pull from Galilean moons (mainly by Io) creates tidal flow and perturbed tidal bulge in the gas giant | Laplacian tidal equation | Unstable (non-conservative) | Idini and Stevenson (2022); Jermyn et al. (2017) |
| 20 | Nonlinear drift waves | Interaction between the solar wind and Jovian dusty plasma results in pressure gradient causing nonlinear drift waves | Nonlinear hydrodynamic equations | Unstable (non-conservative) | Saleem et al. (2012) |

As of now, Jupiter has been visited by a total of nine distinct spacecraft missions (Al-Yousef et al. 2021; Bolton et al. 2015; Connerney et al. 2017; Gurnett et al. 2002; Moslem et al. 2019). The Jupiter exploration mission has started in 1970s, when the *Pioneer 10* and *Pioneer 11* flyby spacecrafts have passed around Jupiter; thereby, detecting its magnetic field. The detectors of these two spacecrafts have recognized only specifically large ($>$ several micrometers) constitutive particles confirming mainly the fluidic composition of the Jupiter. Later, in 1979, the *Voyager 1* and *Voyager 2* flyby spacecrafts have revealed some mysterious observations of Jupiter's moons, rings, magnetic field, and the active volcanoes on Io. It has vastly improved our knowledge on dust dynamics in the Jovian system. In 1992, the *Ulysses* spacecraft has flown by the Jupiter and investigated the Jovian dust and the magnetosphere (Bolton et al. 2015; Moslem et al. 2019; Smith et al. 1992). In 1995, for the first time, the *Galileo* orbiter has gone into the orbit around Jupiter and gathered a large amount of information about the Jovian system. It has observed the complex plasma interactions in Io's atmosphere creating immense electrical currents coupling to Jovian magnetosphere (Bolton et al. 2015; Gurnett et al. 2002).

In 2000, the *Cassini* probe has provided some of the highest-resolution images ever taken of the Jupiter while flying by it. Several different types of radio emissions have been detected by the *Cassini* probe during its approch to Jupiter (Gurnett et al. 2002). The immense presence of the positively charged dust particulates in the Jovian magnetosphere has been confirmed by this probe (Bolton et al. 2015; Gurnett et al. 2002; Moslem et al. 2019). Later in 2007, the *New Horizons* spacecraft has flown by Jupiter for a gravity assist on its way to Pluto. It has made refined measurements of the orbits of Jupiter's inner moons. The probe has measured the structure inside the eruptions on Io, studied all four Galilean moons in detail, and made long-distance studies of the outer moons Himalia and Elara. This spacecraft also has studied Jupiter's Little Red Spot, Jupiter's magnetosphere, and its faint ring system (Bolton et al. 2015).

Most recently the *Juno* orbiter spacecraft has entered a polar orbit of Jupiter in 2016 (Bolton et al. 2015; Connerney et al. 2017; Saur et al. 2018). The spacecraft is studying the planet's interior structure, atmospheric composition, gravity and magnetic field, Jovian lightning, polar magnetosphere, atmospheric dynamics, and intense aurorae. *Juno* is also gathering clues about Jupiter's evolution, including whether there exists a rocky core, the amount of water present within the deep atmosphere, and how the mass is distributed within the planet. In this direction, *Juno* also studies Jupiter's deep supersonic winds ($\sim 600$ km h$^{-1}$ m$= 166.6$ s$^{-1}$) (Bolton et al. 2015). *Juno*'s wave detecting instrument has observed different radio- and plasma-wave phenomena in the polar magnetosphere during its first perijove pass (Connerney et al. 2017).

The whole Jovian magnetosphere system is nourished by plasma constituents which are mainly produced inside the magnetosphere with some external sources (Al-Yousef et al. 2021; Bolton et al. 2015; Saleem et al. 2012; Saur et al. 2018; Smith et al. 1992). Jupiter's moon Io, the most volcanically active body in our solar system, plays as the strongest internal source along with the minor contributions from the Europa and other Galilean moons as well as the Jovian ionosphere (Al-Yousef et al. 2021; Bolton et al. 2015; Saur et al. 2018; Tsuchiya et al. 2018). The high energetic supersonic solar wind and galactic cosmic rays are the main external plasma sources that feed the Jovian magnetosphere (Al-Yousef et al. 2021; Bolton et al. 2015).

The observations made by in situ wave detection instruments of different spacecrafts have provided direct evidences in support of the presence of diversified plasma waves in the Jovian magnetosphere (Al-Yousef et al. 2021; Bolton et al. 2015; Connerney et al. 2017; Gurnett et al. 2002; Saleem et al. 2012; Saur et al. 2018; Smith et al. 1992; Tolba et al. 2021). The Jovian magnetosphere is drastically influenced by continuous and strong ejection of neutral particles and plasmas from the four Galilean moons (Io, Europa, Ganymede, and Callisto) embedded deep inside the magnetosphere as well as the supersonic solar wind (Al-Yousef et al. 2021; Bolton et al. 2015; Saur et al. 2018; Tsuchiya et al. 2018). As a result, different types of collective plasma wave structures are formed. Such excited wave structures play important roles to accelerate the energetic charged particles in the magnetosphere yielding different mystic plasma-based activities, such as the Jovian auroras, Alfvén waves, Jovian hectometric radio emission and auroral extreme ultraviolet emissions triggered by interplanetary shocks, etc. (Bolton et al. 2015; Connerney et al. 2017; Gurnett et al. 2002; Saleem et al. 2012; Saur et al. 2018). Jupiter rotates very fast about its own axis of rotation (in about $10$ hr $= 3.6 \times 10^4$ s) (Bolton et al. 2015; Smith et al. 1992; Tolba et al. 2021; Tsuchiya et al. 2018). This swift rotation of Jupiter, conjugating with its strong magnetic field could also affect the dusty plasma components of the Jovian magnetosphere creating diversified linear and nonlinear collective modes, such as modified dust-ion-acoustic waves (DIAWs), dust-acoustic waves (DAWs), and so forth (Al-Yousef et al. 2021; Saleem et al. 2012; Tolba et al. 2021). Thus, the Jovian environment has become one of the most fascinating plasma laboratories in our solar system. A thorough idea of the configuration and dynamics of the Jovian magnetosphere could be the key to seeing various similar astroplasmic systems, like Saturn's magnetosphere, Uranus's and Neptune's atmosphere, etc. (Table 2). It could enable us to perceive the mysterious giant exoplanet systems and the inherent plasma-based accretion processes during the final stage of gas giant planets formation in the parent protoplanetary disks (PPDs), and so forth (Chen and Bai 2022).

Table 2

Chronological Jupiter exploration with spacecrafts

| Spacecraft (Year) | Type | Activity cum discovery |
|---|---|---|
| Pioneer 10 (1973) | Flyby | (a) It has obtained the first close-up images of the planet<br>(b) It has charted the Jupiter's intense radiation belts and probed the Jovian inherent magnetic field<br>(c) It has discovered the liquid composition of Jupiter predominantly<br>(d) It has transmitted data on the magnetic fields, energetic particle radiation and dust populations in interplanetary space |
| Pioneer 11 (1974) | Flyby | (a) It has obtained some dramatic images of the Great Red Spot<br>(b) It has made the first observation of the immense polar regions<br>(c) It has taken about 200 images of the moons of Jupiter and determined the mass of Jupiter's moon Callisto<br>(d) It has indicated that the Jovian magnetosphere changes its boundaries as it is buffeted by the solar wind |
| Voyager 1 (1979) | Flyby | (a) It has discovered two Jovian mons: Thebe and Metis<br>(b) It has found out ongoing volcanic activity on the moon Io<br>(c) It has also discovered that the Jupiter is surrounded by a thin ring<br>(d) It has portrayed 10 rotations of Jupiter |
| Voyager 2 (1979) | Flyby | (a) It has revealed that the Jovian Great Red Spot is a complex storm moving in a counter-clockwise direction<br>(b) It has confirmed the existence of a thin circumvent ring<br>(c) It has studied the thermal features of the giant planet<br>(d) It has also discovered a previously unknown moon, later named Adrastea, orbiting Jupiter just outside its rings |
| Ulysses (1992) | Flyby | (a) It has gathered data about Jupiter's magnetosphere<br>(b) It has studied the planet's plentiful source of radio waves<br>(c) It has studied the streams of energetic charged particles and dust<br>(d) It has shown the solar wind coupling with Jovian magnetosphere |
| Galileo (1995–2003) | Orbiter | (a) It has gathered evidence for the existence of a saltwater ocean beneath the Jovian moon Europa's icy surface<br>(b) It has discovered extensive volcanic processes on the moon Io<br>(c) It has also found that the giant moon is the first moon known to possesses its own magnetic field<br>(d) It has probed Jupiter's thunderstorms (larger than the Earth's) |
| Cassini (2000) | Flyby | (a) It has captured about 26,000 images of Jupiter and its moons creating the most detailed global portrait of Jupiter yet<br>(b) The images have revealed a dark oval around 60 degrees north latitude, which was a giant storm like the Great Red Spot<br>(c) It has made the first thorough mapping of Jupiter's temperature and atmospheric composition<br>(d) This probe has confirmed the presence of the positive dust charge in the Jovian magnetosphere |
| New Horizons (2007) | Flyby | (a) It has observed lightning near the planet's poles, the life cycle of fresh ammonia clouds, boulder-size clumps speeding through the planet's tenuous rings, the structure inside volcanic eruptions on its moon Io, and the path of charged particles traversing the planet's long magnetic tail<br>(b) It has observed heat-induced lightning strikes in the polar regions the first polar lighting ever observed beyond Earth<br>(c) It has studied the violent storm activity in Jupiter<br>(d) It snapped the first close-up images of the Little Red Spot |
| Juno (2016) | Orbiter | (a) Juno has gathered information about Jovian lightning<br>(b) It has provided the first views of Jupiter's north pole, as well as provided the insight about Jupiter's aurorae<br>(c) It has found a change in the gas giant's magnetic field during its time spent in the orbit<br>(d) It is observing Jupiter's gravity and magnetic field, atmospheric dynamics and composition, and evolution |

This manuscript is organized in a standard pattern layout as follows. Apart from the introduction part as already presented in Sects. 1, 2 gives a brief overview and broad literature survey about the thematic problem of Jovian magnetospheric stability. In Sect. 3, the relevant thermo-statistical distribution law is discussed alongside microphysical aspects. Section 4 is devoted to describe the current objectives of topical research of growing explorative interest in the concerned direction. Finally, the main conclusions drawn from our overall study and survey along with highlighted open future scope are summarily presented in Sect. 5.

## 2. A brief overview

The collective convective circulation dynamo action of the charged particles constituting the interior of the Jupiter produces the intense surface magnetic field of the Jovian system. This strong magnetic field coupled with Jupiter's fast swift rotation creates a unique magnetosphere with plethora of linear and nonlinear collective wave structures (Al-Yousef et al. 2021; Bolton et al. 2015; Saleem et al. 2012; Tolba et al. 2021). This strong magnetic field of the Jovian magnetosphere impedes the motion of charged particles moving perpendicular to the magnetic field and accelerates these particles to move parallel to the magnetic field (Moslem et al. 2019). In the presence of dust grains, some of the electrons (ions) attach with dust to form negatively (positively) charged dust grains and some remaining are bounded back and forth in the potential well by losing energy continuously. The energetic electrons (ions) hereby may not follow the classical thermal Maxwellian-Boltzmann distribution and are trapped to the nonlinear resonant interaction of these particles with positive (negative) potentials (Ahmad et al. 2013; Amin et al. 2010; Annou et al. 2015; Arab et al. 2020; Bolton et al. 2015; Bouziane and Annou 2021; Connerney et al. 2017; Divine and Garrett 1983; Enya et al. 2022; Roussos et al. 2022 ). Thus, we see that the Jovian magnetosphere is an important planetary space regime where highly energetic charged particles, such as electrons, protons, and heavier ions, are trapped significantly. Accordingly, the hot electrons (ions) in the Jovian plasmas follow the trapping-vortex-like distribution dictated by the Cairns–Gurevich (CG) thermostatistical law (Amin et al. 2010; Annou et al. 2015; Arab et al. 2020; Bouziane and Annou 2021). As the potential well for the electrons (ions) is not the potential well for being trapped by the ions (electrons), the distribution considering the trapping of both species simultaneously cannot be physically possible (Kassem et al. 2022; Popel et al. 2003; Gurevich 1968). The origin of the CG distribution function is bi-fold in thermostatistical nature in space and astrophysical circumstances. Accordingly, it is modeled with the help of two distinct non-isothermal behavioural features. The first one is the Cairns distribution function introduced by Cairns (Cairns et al. 1996, 1995), where the non-isothermality is introduced through the background non-uniform ionic distribution. In other words, there always exist diversified zeroth-order gradient (inhomoginity) forces even in the equilibrium configuration in such plasma situation. The second aspect is the Gurevich distribution function prescribed by Gurevich (Gurevich 1968) describing the adiabatic trapping of the lighter plasma constitutive particles in the plasma potential well depending on the corresponding polarity effects (Fig. 1).

**Fig. 1**

A schematic sketch of the multi-layered cross-sectional view of Jupiter

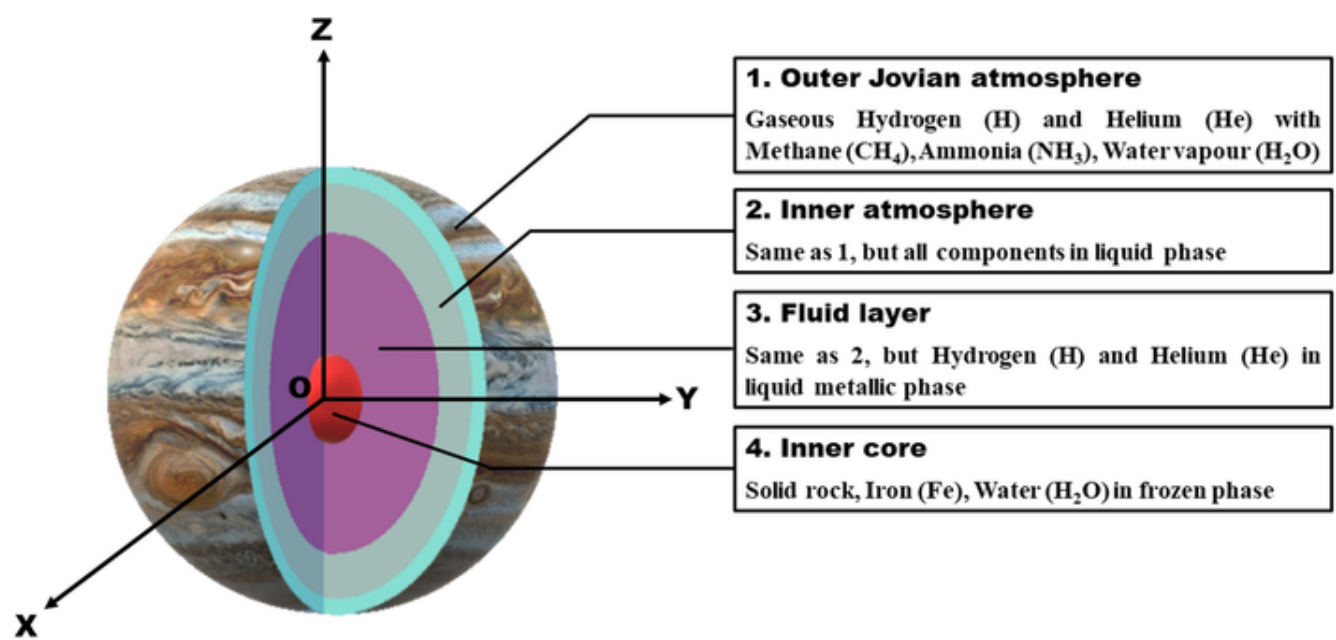

The distance between the Sun and the Jupiter is quite far ($\sim$ 5 AU). Hence, the solar wind becomes rarefied at the Jovian space–time scales with the particle density $2.2 \times 10^5$ m$^{-3}$, mean speed $6 \times 10^5$ m s$^{-1}$ and mean temperature $\sim 3 \times 10^4$ K (Al-Yousef et al. 2021). The streaming supersonic energetic solar wind electrons and protons interact with the Jovian dusty plasma components succeeded by various collective processes including the **AQ2** photoionization, dust charging, secondary electron emission, charged particle collisions, and absorption or emission of radiation yielding rise of various linear and nonlinear waves (Al-Yousef et al. 2021; Saleem et al. 2012). The observations by the *Cassini* spacecraft have confirmed the abundant presence of positively charged dust grains in the Jovian magnetosphere (Bolton et al. 2015; Gurnett et al. 2002; Moslem et al. 2019). This type of charging process usually takes place due to the photoemission caused by flux of ultraviolet (UV) photons, thermionic emission induced by radiative heating, and the secondary electron emission from the dust grain surface triggered by intense radiations from nearby stellar objects (Sun-like) in such space plasma environments (Ahmad et al. 2013; Al-Yousef et al. 2021; Amin et al. 2010; El-Taibany et al. 2022; Moslem et al. 2019). A considerable number of extensive investigations have been reported recently speculating different kinds of linear and nonlinear wave structures subsistent in the Jupiter's magnetospheric region resulting actively from the dynamic collective interactions between the high energetic supersonic solar wind particles pervading the interplanetary space and the Jovian dusty plasmas (Al-Yousef et al. 2021; Saleem et al. 2012; Saur et al. 2018; Tolba et al. 2021).

The dynamics of trapped plasma particles in a dusty plasma system has been an interesting fact to study since last few decades. The effects of plasma particle trapping on dust-acoustic solitary waves in an opposite polarity dusty plasma medium (with bipolar dust) have been reported (Ahmad et al. 2013). When both positively and negatively charged dust grains exist in a plasma system, the plasma medium is termed as "opposite polarity dusty plasma" or "bipolar dusty plasma" (El-Taibany et al. 2022). Deriving a pseudo-potential, a theoretical investigation has been carried out therein on dust-acoustic solitary waves in a complex plasma system composed of adiabatic positively and negatively charged dust grains, non-isothermal electrons and ions (following vortex-like distribution) (El-Taibany et al. 2022). Employing the reductive perturbation method, the propagation of small finite amplitude solitary structures has been theoretically examined. It has been interestingly found therein that, in such an opposite polarity dusty plasma medium, the conjoint effects of dust polarity, trapping of the constitutive plasma particles (electrons and ions), and dust fluid temperatures significantly exhibit diverse impacts on the basic features of the dust-acoustic solitary wave structures (El-Taibany et al. 2022).

Later on, a study of an electrostatic asymmetric shielding potential as well as dynamical oscillatory plasma wakefield potential in an unmagnetized plasma has been reported (Moslem et al. 2019). The potential of a moving test charge interacting with the solar wind in dusty magnetosphere of Jupiter has been derived and investigated both analytically and numerically (Moslem et al. 2019). This reported model has been applied to Jovian magnetosphere that contains positively charged dust grains, Maxwellian electrons and ions, as well as streaming energetic solar wind electrons and protons. Using algebraic manipulations, the Debye–Hückel and the Wakefield potentials have been simply derived. It has been shown therein that the normalized Debye potential decreases exponentially with the variation in axial distance while the normalized wakefield potential is strongly influenced by the plasma temperatures and the velocity of solar wind electrons and protons. It has also been observed that both the background ion and electron temperatures ($T_{e,i}$) exhibit the same qualitative behavior on the wakefield profile. It is interestingly seen from the report that for dusty plasma, increasing both $T_i$ and $T_e$ lead to decrease the wakefield potential amplitude (Moslem et al. 2019). Moreover, the wakefield potential exhibits oscillatory behavior with increasing $\xi$-value and for higher values of $T_{i,e}$. However, for dust-in-plasma case, an increase in both $T_i$ and $T_e$ can lead to an enhancement of the wakefield potential amplitude, and vice versa. Furthermore, the increasing streaming speed of solar wind electrons and protons decreases both the potential amplitude and the wave frequency in both cases of dusty plasma and dust-in-plasma (Moslem et al. 2019).

In recent past, for the first time, a theoretical investigation has been reported considering the effects of the trapped nonthermal polarization force called the Cairns–Gurevich (CG) polarization force on dust-acoustic solitons in collisionless dusty plasmas (Arab et al. 2020). The dusty plasma has been assumed therein to be consisted of thermal Maxwellian electrons, negatively charged dust grains, and trapped non-isothermal ions. A three-dimensional Cairns–Gurevich (CG) distribution has been defined therein to describe the evolution of the energetic ions as well as those nonthermal ions trapped in the plasma negative potential well (Arab et al. 2020). Using this CG distribution function, the density and polarization force expressions have been derived accordingly. This report has revealed that the effects of the polarization force are significantly modified due to the presence of the trapped nonthermal ions (Arab et al. 2020). It has been observed therein that the magnitude of the polarization force decreases with an enhancement in the degree of ionic nonthermality $\boxed{\alpha}$. Moreover, in that report, it has been speculated that an increament in the nonthermal character of ions following the CG distribution leads to a growth in the amplitude and width of DA waves, and vice versa (Arab et al. 2020).

Recently, an investigation of the propagation of arbitrary amplitude dust-acoustic solitary waves in the complex unmagnetized plasma system of the Jovian atmosphere for distance greater than 15 $\boxed{R_J}$ has been reported (Al-Yousef et al. 2021). At this region of Jovian magnetosphere $\boxed{(15R_J < r < 60R_J)}$, the magnetic field stretches radially significantly. The heavy amount of plasma stream from Io and the centrifugal force feature of rapidly rotating Jovian plasma environment result in such type of radial stretching of the magnetosphere closed to the Jovian equatorial plane (Bolton et al. 2015). The hydrodynamic Jovian plasma model reported therein has been considered to be composed of positively charged dust grains, Maxwellian electrons, and ions those interact with solar wind protons and electrons (Al-Yousef et al. 2021). Therein, the basic equations describing the plasma system have been reduced to an energy-balance-like equation containing Sagdeev potential which describes the electrostatic potential of the system. By performing constructive numerical analysis and solving the Sagdeev potential, it has been found that though subsonic and supersonic acoustic waves both can exist, the dominant pulses are seen to be in supersonic modes and the pulses have positive potential. Moreover, it has been shown that the dust number density, solar wind proton number density, temperature, and streaming speed are the main parameters to change the soliton existence region. It has been found therein interestingly that, as the dust grain number density increases, the width of the solitons decreases, while, enhancing the amplitude (Al-Yousef et al. 2021).

Very recently, considering a generalized hydrodynamic jovian dusty plasma model, a study of arbitrary amplitude dust-acoustic waves in Jovian magnetosphere has been reported in the framework of the Sagdeev pseudo-potential method (Tolba et al. 2021). The adopted dusty plasma model has been assumed therein to be homogeneous, collisionless, and magnetized consisting of positively charged dust grains along with background thermal elctrons and ions. The dust grains have been described therein via fluid equations while the electrons and ions have been assumed to follow the Boltzmann distribution (Table 3). An energy-balance-like evolution eqution, revealing the law of conservation of energy in light of the Sagdeev energy integral techniques, has been derived and analyzed numerically (Tolba et al. 2021). It has been found therein that the reported data of the obsereved physical parameters produce only positive potential nonlinear DA cnoidal waves without a company of any other nonlinear waves (Tolba et al. 2021). The effects of the external magnetic field, Mach number, and directional cosine parameters have been studied and deeply criticized. Moreover, it has been found therein that the amplitude and width of positive potential DA cnoidal waves increase with an increment in the external magnetic field (B) and the Mach number (M), respectively (Tolba et al. 2021).

Table 3

A glimpse of recent studies

| Source | Description | Limitation |
|---|---|---|
| Ahmad et al. (2013) | (a) Plasma particle trapping effects on dust-acoustic solitary waves in opposite polarity dusty plasma have been studied<br>(b) Adopted complex plasma system is composed of adiabatic positive and negative dust, non-isothermal electrons and ions (with vortex-like distributions)<br>(c) Nonthermal effects in the form of either trapped electrons or ions on the dust-acoustic solitary waves have been studied<br>(d) Propagation of finite amplitude solitary structures has been seen perturbatively<br>(e) Conjoint effects of dust polarity, trapping of plasma particles, and dust grain temperature significantly modify the basic features of the nonlinear DAWs | (a) CG distribution has not been taken into consideration<br>(b) Positive and negative dust grains have been considered<br>(c) Non-isothermal distribution of dust has not been taken |
| Tolba et al. (2015) | (a) Nonlinear rogue waves evolution in Jovian plasma composed of positive dust grains, streaming electrons and positive ions, as well as Maxwellian electrons and positive ions has been investigated<br>(b) Using a perturbation method, the basic set of fluid equations is reduced to a nonlinear Schrödinger equation (NLSE)<br>(c) Existence region of rogue waves depends mainly on the dust-acoustic speed, temperature ratio, and the streaming densities of the ions and electrons<br>(d) Supersonic rogue waves are much taller than the subsonic one by $\sim 25$ time | (a) Trapping of electrons and ions has not been addressed<br>(b) CG distribution has not been taken into consideration<br>(c) Thermal Maxwell–Boltzmann distribution of the electrons and ions has been considered<br>(d) Non-isothermal distribution of charged dust particulates has not been considered |
| Tolba et al. (2017) | (a) Dust-acoustic nonlinear periodic (cnoidal) waves are studied in positively charged dusty plasmas (Jupiter-like) interacting with streaming solar wind<br>(b) Using the reductive perturbation method, nonlinear evolution equations are derived<br>(c) Cnoidal wave solutions are obtained by the Sagdeev pseudopotential analysis<br>(d) Amplitude and wavelength of the negative cnoidal waves are much shorter than the positive cnoidal waves<br>(e) Cnoidal pulses cannot propagate for supersonic dust-acoustic phase speed, but can exist for subsonic phase speed only | (a) Trapping of electrons and ions has not been considered<br>(b) CG distribution has not been taken into consideration<br>(c) Maxwell–Boltzmann law for the background electrons and ions has been considered<br>(d) Non-isothermal distribution of charged dust particulates has not been addressed |
| Moslem et al. (2019) | (a) Potential of a moving test charge interacting with the solar wind in dusty magnetosphere of Jupiter has been investigated analytically and numerically<br>(b) Jovian magnetosphere here contains positive dust particulates, Maxwellian electrons and ions, and streaming solar wind electrons and protons<br>(c) Debye–Hückel and Wakefield potentials have been simply derived and analyzed | (a) Thermal Maxwell–Boltzmann distribution of the electrons and ions has been considered<br>(b) Trapping of electrons and ions has not been considered<br>(c) Non-isothermal distribution of dust grains has not been taken into consideration |
| Arab et al. (2020) | (a) Theoretic study has been reported with the effects of trapped nonthermal CG polarization on dust-acoustic solitons in collision-less dusty plasmas<br>(b) Dusty plasma considered herein consists of Maxwellian electrons, negatively charged dust grains, and trapped ions<br>(c) Three-dimensional CG distribution describing the evolution of the trapped ions has been derived to study the density and polarization force on acoustic waves | (a) CG distribution for only ions has been taken into account<br>(b) Ion trapping is studied, trapping of electrons has not been considered<br>(c) Dust grains considered here are negatively charged<br>(d) Only linear perturbation method has been carried out<br>(e) Non-isothermal behavior of dust has not been considered |

| Source | Description | Limitation |
|---|---|---|
| Al-Yousef et al. (2021) | (a) Propagation of arbitrary amplitude dust-acoustic solitary waves in unmagnetized Jovian dusty plasma for distance greater than 15 $R_J$ has been investigated therein<br>(b) Positive dust grains, thermal Maxwellian electrons and ions has been considered as the main constituting particles<br>(c) Energy-balance-like equation containing Sagdeev pseudo-potential has been derived and numerically analyzed | (a) Maxwellian thermal electrons and ions has been considered<br>(b) Trapping-vortex-like distribution (CG distribution) of plasma particles has not been taken into account<br>(c) Nonthermal distribution of dust has not been considered |
| Tolba et al. (2021) | (a) Generalized hydrodynamic plasma model composed of positively charged dust grains, Maxwellian ions and electrons has been considered to study arbitrary amplitude dust-acoustic waves<br>(b) Assumed dusty plasma is homogeneous, collision-less, and magnetized<br>(c) Evolution equation containing Sagdeev pseudo-potential has been derived and numerically analyzed<br>(d) Reported data produces only nonlinear DA cnoidal waves (with positive potential only) without a company of any other nonlinear waves<br>(e) Effect of the external magnetic field, Mach number, and directional cosine parameters has been exhaustively studied<br>(f) Amplitude and width of positive potential DA cnoidal waves increase with an increment in the external magnetic field (B) and Mach number (M), respectively | (a) Maxwellian thermal electrons and ions has been considered<br>(b) Trapping of electrons and protons has not been taken<br>(c) Trapping-vortex-like CG distribution of plasma particles has not been taken<br>(d) Non-isothermal distribution of dust has not been taken |

# 3. **Relevant thermostatistics**

A numerous studies and analyses on linear and nonlinear dust-acoustic waves have reported that a significant amount of electrons and ions are trapped in positive and negative wave potential, respectively (Ahmad et al. 2013; Amin et al. 2010; Annou et al. 2015; Arab et al. 2020; Bouziane and Annou 2021). The corresponding modified type of thermostatistical distribution law relevant here clearly exhibits a deviation from the thermal Maxwell–Boltzmann distribution of the electron and ion population densities, thereby introducing a new trapping-vortex-like electron and ion distributions in the defined phase space (Ahmad et al. 2013). Such distributions of lighter plasma particles (electrons and ions), indeed, modify the behavior of collective coherent (nonlinear) structures, such as acoustic waves, solitons, shocks, etc., against the classical thermal Maxwell–Boltzmann scenario (Ahmad et al. 2013; Amin et al. 2010).

The inherent strong magnetic field of the Jupiter's magnetosphere obstructs the motion of charged particles moving perpendicular to the magnetic field, hence accelerating the charges to move parallel to the magnetic field (Moslem et al. 2019). Some of the electrons (ions) attached with the Jovian dust to form negatively (positively) charged dust grains and some remaining are bounded back and forth in the potential well by losing energy continuously. Accordingly, The energetic electrons (ions) are trapped to the nonlinear resonant interaction potentials of these particles with positive (negative) polarities. As a consequence, local thermodynamical equilibrium configuration is disturbed and a nonthermal anti-equilibrium configuration is reorganized. So, the constitutive lighter species do not follow the classical thermal Maxwellian-Boltzmann distribution law (Ahmad et al. 2013; Annou et al. 2015; Arab et al. 2020; Bolton et al. 2015; Bouziane and Annou 2021; Connerney et al. 2017; Enya et al. 2022). To model such type of trapping-vortex-like distribution of hot electrons (ions), a new distribution law of nonthermal non-energetic trapped plasma constituents has been introduced for studying the nonlinear dust-acoustic wave structures (Annou et al. 2015; Arab et al. 2020; Bouziane and Annou 2021). This non-thermal thermo-statistical distribution law, called the "Cairns–Gurevich (CG) distribution", describes simultaneously the evolution of the energetic free electrons (ions) alongside the non-energetic trapped electrons (ions) in the positive (negative) plasma potential well (Annou et al. 2015; Arab et al. 2020; Bansal and Gill 2022; Bouziane and Annou 2021).

The CG distribution law for the electron population density with all the customary notations of the relevant physical variables (Annou et al. 2015; Bouziane and Annou 2021) is cast as

$$n_e = 1 + (1 - \beta_e)\,\phi - \frac{4\sqrt{\pi}}{3}\phi^{3/2} + \left(\frac{1}{2} + \beta_e\right)\phi^2; \tag{1}$$

It may be seen that the terms involving the $\phi$-integral powers correspond to the free electrons; while, the half-integral powers designate the trapped electrons in the plasma system.

The CG distribution law for the constitutive ions in generic notations is similarly given as (Arab et al. 2020; Popel et al. 2003)

$$n_i = n_{i0}\left[C_1 \exp\left(-\frac{e\phi}{T_i}\right)\text{erfc}\left(\sqrt{-\frac{e\phi}{T_i}}\right) + C_2\sqrt{-\frac{e\phi}{\pi T_i}}\right]; \tag{2}$$

where,

$n_{e(i)}$: Electron (Ion) number density;

$\phi$: Electrostatic potential;

$$C_1 := 1 + \beta_i\left(\frac{e\phi}{T_i}\right) + \beta_i\left(\frac{e\phi}{T_i}\right)^2;$$

$$C_2 := 2 + \left[3\beta_i + \frac{4}{3(1+3\alpha_i)}\right]\left(\frac{e\phi}{T_i}\right) + \frac{56}{15}\beta_i\left(\frac{e\phi}{T_i}\right)^2;$$

$\alpha_{e(i)}$: Degree of electronic (ionic) non-thermality (depicting the degree of deviation from the Maxwell–Boltzmann distribution law);

$\beta_{e(i)} = \frac{4\alpha_{e(i)}}{1+3\alpha_{e(i)}}$: Electron (Ion) non-thermality parameter;

$\text{erfc}(x) = [1 - \text{erf}(x)] = \frac{2}{\sqrt{\pi}}\int_x^{\infty} e^{-y^2}\,dy$: Complementary error function.

It may be pertinent to add here that, in Eq. (2), the first term corresponds to the constitutive free ions; while, the trapped ions are well represented by the second term in the considered plasma system.

# 4. **Current objectives**

The strong plasma outflow from the Io and others Galilean moons as well as the solar wind immensely interacts with Jupiter's large intrinsic magnetosphere, its constitutive positively charged dust grains, and the other plasma constituents (Al-Yousef et al. 2021; Bolton et al. 2015; Moslem et al. 2019; Saur et al. 2018; Tolba et al. 2021). It results in different collective instability processes and propagation dynamics of diversified linear and nonlinear plasma wave structures in the Jovian atmosphere, which are yet to be well understood (Al-Yousef et al. 2021; Bolton et al. 2015; Moslem et al. 2019; Tolba et al. 2021). The detailed thermostatistical description law of the trapping mechanism of both the lighter species (electrons + ions) against the heavier positively charged dust particulates (microspheres) is still lying as an open challenge. In our planned meta-analysis, we are applying a hydrodynamic model formalism on the Jupiter's nonthermal dusty magnetosphere to investigate the dust-acoustic wave (DAW) dynamics in the presence of justifeable and realistic novelties with high applicability in such astronomic and space plasma environments.

To carry out a semi-analytical study of the refined DAW dynamics in the nonthermal dusty plasma system mimicking Jupiter's magnetosphere, we are planning to use nonthermal vortex-like Cairns–Gurevich (CG) distribution law (Annou et al. 2015; Arab et al. 2020; Bouziane and Annou 2021). It would describe simultaneously the evolution of the energetic free electrons (ions) along with those are trapped (non-energetic) in the plasma potential well. In other words, it would consider the combine effect of energetic and trapped (non-energetic) electrons (ions) in the medium (Annou et al. 2015; Arab et al. 2020; Bouziane and Annou 2021). To concretize our proposed theoretical investigations yet to be executed in such thermo-statistical Jovian plasmas, we want to model our foremost future objectives strategically as:

(i) It would be very interesting and useful to see the linear dynamics of the bulk-acoustic waves (DAWs) in the Jovian magnetosphere with trapped non-energetic nonthermal electrons (ions) dictated by the CG distribution and free energetic ions (electrons) along with the positively charged dust particulates in an inhomogeneous and non-uniform configuration (Annou et al. 2015; Arab et al. 2020; Bouziane and Annou 2021).

(ii) It is still an open challenge to study the weakly nonlinear dynamics of the DAWs in the above model using the standard method of multiple scaling techniques in the presence of diversified realistic complications, such as the dust-mass distribution, dust-charge distribution, and so forth (Ata-Ur-Rahman et al. 2019; Bansal and Gill 2022).

(iii) A relevant analysis could be carried out about the strongly nonlinear dynamics of the DAWs in the same model using the Sagdeev pseudo-potential techniques founded on the universal energy conservation principle (Ahmad et al. 2013; El-Taibany et al. 2022; Tolba et al. 2021).

(iv) An extension study can be relevantly conducted to explore strongly nonlinear DAW dynamics using the modified Sagdeev pseudo-potential techniques in the presence of diversified dissipative factors (Akbari-Moghanjoughi 2017; Ali Shan et al. 2019).

(v) Investigating the effects of diversified thermo-statistical distribution laws on the collective wave excitation processes is indeed believed to be a sensible comparative study in the realm of planetary science yet to be well understood (Gupta et al. 2022).

(vi) A remarkable goal may be set here to explore the associated mysterious collective instabilities amid non-uniform equilibrium configurations in the framework of eigenvalue treatments and WKB approaches with variable temperature distribution (Gupta et al. 2022).

(vii) Further, an analysis of the collective wave excitation processes induced by dust-charge variational dynamics could be an important goal yet to be materialized (Amin et al. 2010; Bansal and Gill 2022).

(viii) Lastly, it is planned to revisit Jovian plasma models with a realistic form of equation of state in the presence of poly-tropic moderation effects of thermo-statistical origin (Chen and Bai 2022; Idini and Stevenson 2022).

(ix) As the dust grain charge fluctuates, a full investigation of the DCW dynamics would be an important study in this direction of major plasma concern (Rao 1999, 2000).

(x) In the strongly coupled dusty plasma regime, the study of the linear and non-linear DLWs would find considerable interest in the domain of such Jovian dusty plasmas (Rao 1999, 2000; Shukla and Mamun 2001); and so forth.

We can finally anticipate herewith that our futuristic theoretic study could tentitively bridge a better correlation and consistency with the already reported theoretical and observational predictions availilable in the literature with a significant degree of overall reliability (Al-Yousef et al. 2021; Saleem et al. 2012; Saur et al. 2018; Tolba et al. 2021). Moreover, it could justifiably enable us for further analysis and refinement paving the way of futuristic investigation and realistic applicability in the emerging Jovian plasma stability direction of future space plasma interest bearing non-zero socio-economically applied value, and so forth. It is worth mentioning here that, what is briefly indicated above, is a semi-analytic approach only. A regorous analysis of the collective kinetic structure formation in the corroborative light of diversified multi-satellite detection results would be significantly a more interesting explorative achievement.

# 5. **Conclusion**

We herein report an explorative analytic study compiling the dynamics of dusty plasma environment in the Jupiter's giant magnetosphere. The basic feature of this chronological review lies in the concise description of the collective instabilities and propagation of dust-acoustic waves (DAWs) in the presence of trapped lighter plasma particles along with the heavier positively charged dust grains. The giant Jovian magnetosphere is primarily fed by plasma constituents produced by ionization mechanism of volcanic gasses from the Io with some other minor additive sources. Different spacecraft exploration missions that encountered Jupiter and their precise observations are outlined here chronologically. Furthermore, our pertinent research objectives are summarily proposed herewith in light of sensible validated novelties and tentative future directions. It is noteworthy that many major mysterious facts about the Jovian dusty magnetosphere have still been remaining unsolved for years. The mechanism behind the continuous heating of the plasma moving outwards from the Io's plasma torus, the formation, and the time variation feature of Jupiter's magnetic field, the collective excitation processes in the Jovian magnetosphere, Jovian atomospheric thermostatistical profile, and many other mystical unsolved puzzles is yet to be well modeled and explored systematically.

It is further expected that the ongoing orbiting mission by the National Aeronautics and Space Administration's (NASA's) *Juno* space probe (Connerney et al. 2017 ) and the European Space Agency's (ESA's) proposed *JUpiter ICy moons Explorer* (*JUICE*) mission (2023) (Enya et al. 2022) on Jovian system will shed some light on the unsolved mysteries of the entire planet and its moons (natural satellites). It is strongly anticipated that rigorous observations and explorations on plasma configurations and dynamics of the Jovian dusty magnetosphere system could enlighten us with the understanding of various similar astroplasmic systems as well. It could also be efficacious to explore various microphysical aspects of linear and nonlinear wave structures observed in diversified like space and astrophysical plasma environments. Besides, it especially could aim at investigating extensively the collective instability dynamics excitable in the complex solar wind plasma, Jupiter's magnetosphere, Saturn's magnetosphere, similar giant gaseous exoplanet systems, and so forth. `AQ3`

## Publisher's Note

### Acknowledgements

The authors gratefully acknowledge the active cooperation availed from the Department of Physics, Tezpur University. The dynamic support of the colleagues of Astrophysical Plasma and Nonlinear Dynamics Research Laboratory, Department of Physics, Tezpur University, is duly worth mentioning. The commendable role of the learned anonymous referees in the form of scientific comments and helpful remarks leading to improvement is thankfully appreciated. The financial support received through the SERB Project, Government of India (Grant—EMR/2017/003222), is thankfully recognized.

Data availability

No new data were created or analyzed in this study.

## Declarations

***Conflict of interest***    The authors declare no competing interests.